\documentclass[%
 reprint,
superscriptaddress,
 amsmath,amssymb,
]{revtex4-2}

\usepackage{graphicx}
\usepackage{dcolumn}
\usepackage{bm}
\begin{document}
\title{Design of a Littrow-type diode laser with independent control of cavity length and grating rotation}

\author{Lucia Duca}
\altaffiliation[Corresponding author: ]
{l.duca@inrim.it}
\affiliation{INRiM, Strada delle Cacce 91, 10135 Torino, Italy}
\author{Elia Perego}
\affiliation{INRiM, Strada delle Cacce 91, 10135 Torino, Italy}
\author{Federico Berto}
\affiliation{INRiM, Strada delle Cacce 91, 10135 Torino, Italy}
\affiliation{Politecnico di Torino, Corso Duca degli Abruzzi 24, 10129 Torino, Italy}
\author{Carlo Sias}
\affiliation{INRiM, Strada delle Cacce 91, 10135 Torino, Italy}
\affiliation{LENS, Via Nello Carrara 1, 50019 Sesto Fiorentino, Italy}
\affiliation{Istituto Nazionale di Ottica-CNR, 50019 Sesto Fiorentino, Italy}



\begin{abstract}
We present a novel extended-cavity diode laser (ECDL) based on a modified Littrow configuration. The coarse wavelength adjustment via the rotation of a diffraction grating is decoupled from the fine tuning of the external cavity modes by positioning a piezo transducer behind the diode laser, making the laser robust against misalignment and hysteresis even with long external cavities. Two laser prototypes with external cavities of different lengths were tested with a 780 nm laser diode, and locked to an atomic reference. We observe a mode-hop-free frequency tunability broader than the free spectral range of the external cavity upon changes of its length. 
The design is well suited to atomic and molecular experiments demanding a high level of stability over time.
%
\end{abstract}




\maketitle

\section{Introduction}
Extended-cavity diode lasers (ECDLs) are fundamental and rather inexpensive sources of coherent and frequency-stabilized light. ECDLs are used in many scientific applications, such as atomic physics and metrology  \cite{Wieman91,Ludlow15}, because of their good tunability, single mode operation, frequency stability and mechanical robustness. Moreover, an advantage of ECDLs is the wide range of wavelengths of the laser diodes (LDs) available in the market and their relatively broad spectral profile, well suited for tuning the emission wavelength. The versatility of the active medium and the small number of elements required for realizing an ECDL make it relatively easy to assemble lasers which fulfill typical experimental requirements for atomic and molecular spectroscopy with relatively simple designs.

In the most common design, a diffraction grating in Littrow configuration \cite{RICCI1995} provides both the optical feedback of the external cavity and the wavelength selection. Although very essential and effective, this design is sensitive to mechanical misalignment and hysteresis - a result of the contradicting demands for a rigid construction and fine tunability of the opto-mechanical components used to tune the laser frequency. Consequently, several designs have been developed with improved damping and isolation from the external environment \cite{Cook2012,Kirilov2015,Loh2006}, and with a careful choice of the grating pivotal point \cite{McNicholl1985,Saliba2009}. An alternative to the Littrow-type ECDL is represented by the Littman-Metcalf design, where a mirror positioned after the grating provides the reflection for the external cavity \cite{Littman1978,McNicholl1985}. However, also in this design the coarse and the fine tuning of the laser frequency are realized by acting on the same mechanical component, and the overall laser power is affected by additional losses with respect to the Littrow design.

Another type of ECDL employs interference filters or etalons instead of a diffraction grating \cite{BAILLARD2006} to decouple the optical feedback of the external cavity and the wavelength selection, thereby reducing the sensitivity of the external cavity feedback against misalignment. The drawback of this design is the temperature sensitivity of the optical interference filter and the generally lower output power which is due to the non-negligible transmission losses of the filter, even at the peak wavelength. In order to have a large enough optical feedback for single mode operation, the losses are compensated with a higher reflectivity of the external cavity mirror, resulting in a net lower output power. 
This design is therefore less convenient for situations in which a secondary amplification stage is not possible, for example for blue wavelength ECDLs.

Here we propose and test an alternative ECDL design based on the Littrow configuration which combines the advantages of the interference filter models with the high output power and temperature stability typical of the Littrow design. The main feature of the new design is the decoupling of the fine-tuning of the laser frequency from its coarse-tuning. In fact, while the coarse tuning and the alignment of the optical feedback are achieved by manually rotating a diffraction grating in Littrow configuration, the frequency fine tuning is realized by moving with a piezo transducer (PZT) the laser diode. 
As a result, the laser frequency follows the shift of the external cavity modes that is realized by changing the external cavity length, while the frequency selection of the grating and the beam pointing remain unchanged. This is particularly convenient for long cavity ECDLs \cite{Kolachevsky:11} where instabilities related to feedback alignment and pivotal point's position are more pronounced. 


\section{The design of a 780~\lowercase{nm} prototype}

\begin{figure}[b!]
	\centering
		\includegraphics[width=0.8\linewidth]{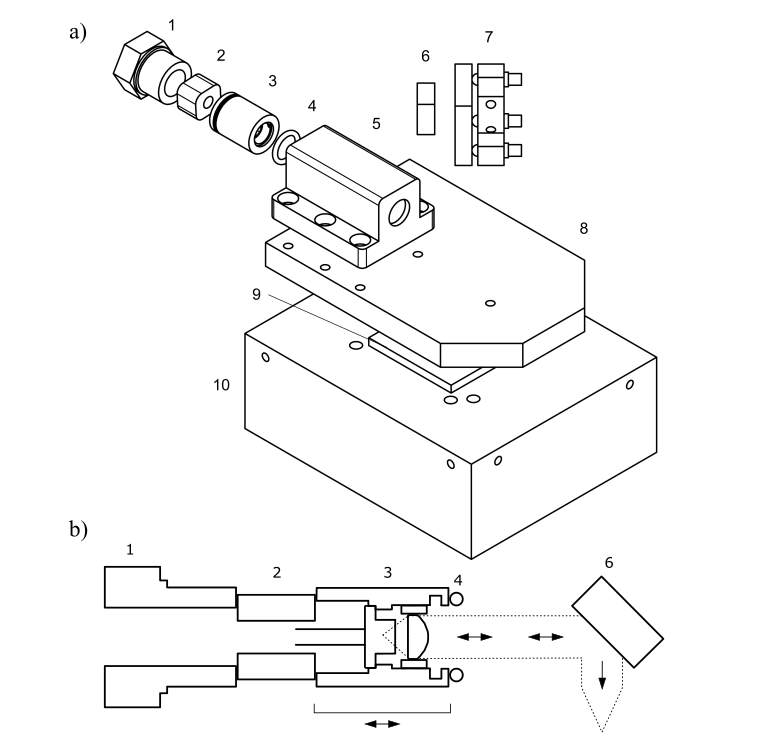}
    \caption{\textbf{Schematic drawing of the ECDL body.} a) The laser diode is mounted in a collimating tube (3) placed between a ring piezo transducer (2) and a rubber o-ring (4). The stack is fixed in a laser head holder (5) by a bushing (1). The grating (6) is placed on an adjustable mirror mount (7) whose angle can be tuned by micrometer screws for coarse wavelength adjustment. The ECDL temperature is stabilized by a Peltier element (9) mounted between the laser base (8) and a thermal sink base (10).
    b) Schematic diagram of the ECDL showing the beam path in dashed line.
    The moving part of the ECDL assembly, i.e. the collimating tube (3), is highlighted. 
}\label{Fig:1}
\end{figure}

Fig.~\ref{Fig:1} illustrates a schematic of the laser prototype that was assembled and tested with an anti-reflection coated 780~nm diode from Eagleyard Photonics (EYP-RWE-0790-04000-0750-SOT01-0000). The diode current is controlled by a commercial laser driver (Thorlabs LDC8002). The external cavity of the laser is created by the rear facet of the diode and a reflective holographic grating with 1800~grooves/mm (Thorlabs GH13-18V, element 6 in Fig.~\ref{Fig:1}) placed in Littrow configuration. The grating also selects one single mode of the external cavity which lies within the frequency envelope of the grating. Although our prototype might look similar to most of the ECDL Littrow designs, it instead has the conceptual difference of having the PZT mounted behind the laser diode and not behind the grating.
This choice effectively decouples the change of external cavity length from the rotation of the grating, corresponding to coarse wavelength selection and feedback alignment. As actuator we use a ring piezo (2) (Piezosystem Jena HPSt 500/15-5/7) placed behind the collimation tube where the laser diode is mounted (3) (Thorlabs LT230P-B, collimating lens: A230TM-B). The cabling for the current control of the diode is passing through the PZT aperture. 
The stack of PZT and laser tube is kept in place and pressed by a bushing (1) against a rubber o-ring (4) that ensures the PZT pre-loading \cite{Sakuta2020}.
The grating is placed on an adjustable mirror mount (7) to fine tune the horizontal and vertical alignment of the first diffraction order back into the diode. 
The ECDL mechanics is mounted on a base (8). A Peltier element (Adaptive ET-161-12-10-E) placed on a thermal sink (10) keeps the temperature of the ECDL stable to within a few millikelvins by using an electronic controller (Thorlabs TED8020). 
All the mechanical parts of the ECDL are machined in an aluminium alloy of series 6000, characterized by good elasticity under mechanical stress.



\section{Characterization of two ECDL prototypes with different cavity lengths}
\subsection{Frequency tunability}

\begin{figure}[b!]
	\centering
		\includegraphics[width=\linewidth]{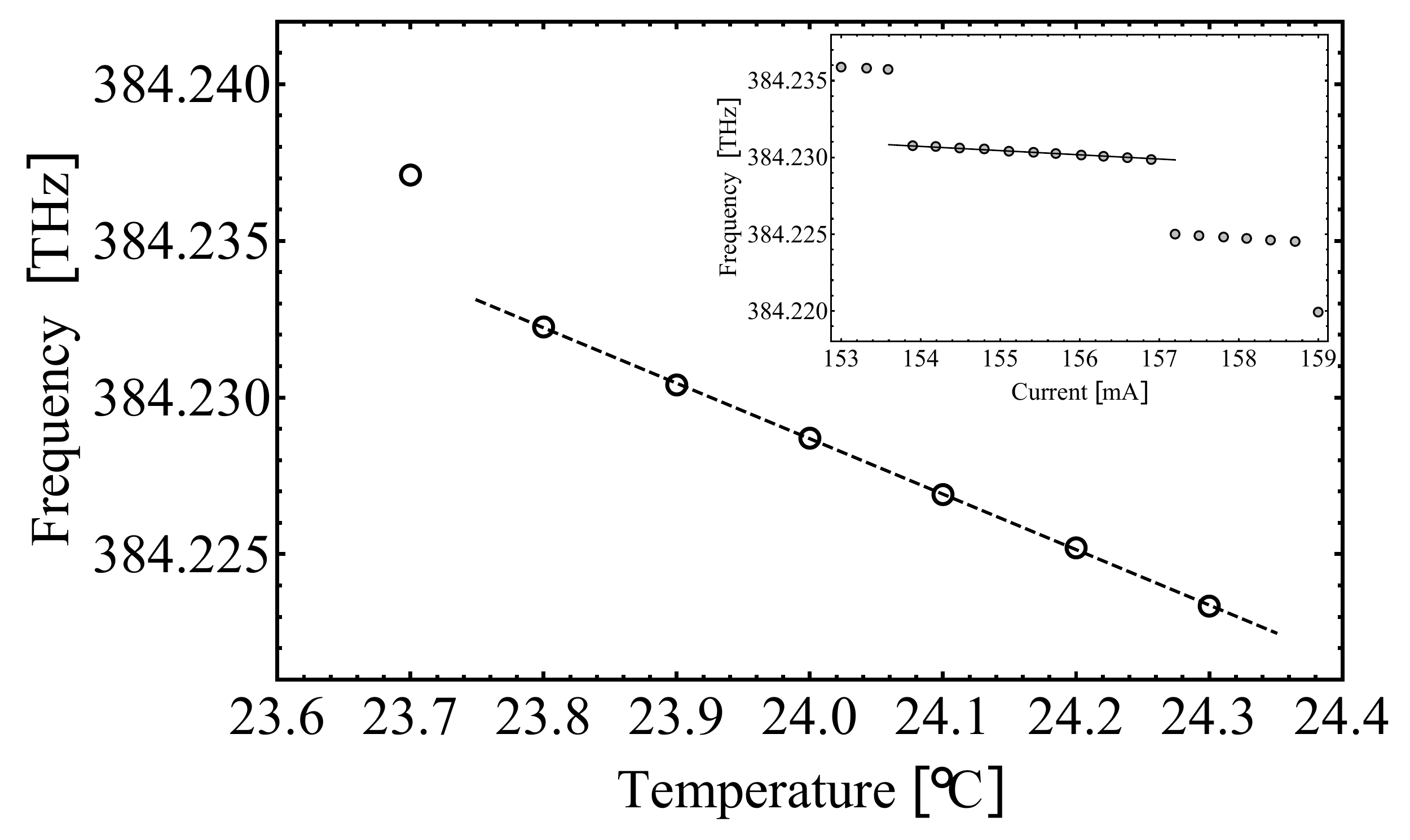}
    \caption{ \textbf{Tuning the prototype's temperature and current.} Tuning the temperature changes the frequency the ECDL as $-~17.9~$GHz/$^\circ$C (empty dots with dashed fit) in the mode-hop free regions. We observed mode-hop free operation for a maximum temperature change of 0.5$^\circ$C. 
    The laser current, instead, finely adjusts the laser frequency (inset). In the stable regions, before the laser jumps to another cavity mode, the frequency changes as $-~274~$MHz/mA. 
    The frequency jumps indicate a FSR of 4.8(2)$~$GHz, which is compatible with the value expected for an external cavity of length L~=~31~mm. 
}\label{Fig:2} 
\end{figure}


The behavior of the laser frequency as a function of the diode temperature and current gives a first estimate of the external cavity FSR. 
The laser frequency can hop between different stability regions by changing the laser temperature - a result of changing both the external cavity length and the spontaneous emission envelope of the diode. As illustrated in Fig.\ \ref{Fig:2}, we observed mode-hop free operation for a maximum temperature change of 0.5$^\circ$C for the laser depicted in Fig.\ \ref{Fig:1}. 
Mode hops are also observed when the diode current is changed, as illustrated in the inset of Fig.~\ref{Fig:2}. From these data we derive that the amplitude of the mode hops is 4.8(2)~GHz, a value that is in agreement with the expected FSRs for a cavity length of about 31 mm. 
The emission frequency is also tuned by changing the length of the external cavity with the PZT. Usually, in a Littrow type ECDL the piezo element is located behind the grating and it changes at the same time the cavity length and the grating angle. There exists an optimal pivotal point for which the cavity resonances and the frequency selected by the grating have the same dependence from the grating angle, ensuring a maximum tunability of the laser frequency~\cite{McNicholl1985, Saliba2009}. However, when this optimal configuration is not realized, e.g. in ECDLs with cavities longer than a few cm, the mode-hop free tunability is limited to a fraction of the free spectral range~\cite{RICCI1995,Kolachevsky:11}. 
In our design, instead, tuning the PZT has the effect of changing the length of the external cavity only. 
In order to test the potentiality of this laser design we assembled two prototypes: one with a relatively short (about 31 mm) and one with a longer (about 130 mm) external cavity. The free spectral ranges (FSRs) of the external cavities are 4.8~GHz and 1.2~GHz respectively. The mechanical mounting system is the one shown in Fig.\ \ref{Fig:1} for the short ECDL. For the long ECDL, instead, we mounted the grating on a separate mirror mount placed externally from the laser base (8) in Fig.\ \ref{Fig:1}. As a consequence, in this prototype the temperature control keeps only the temperature of the diode stable, while the external cavity is free to fluctuate with the room temperature (stable up to 0.5~K). 
The long ECDL is therefore a good test of the laser design when placed in a noisy environment: the more finely spaced the cavity modes are, the more likely it is for a mode hop to occur when the external cavity length is changed.

\begin{figure}[tbp]
	\centering
		\includegraphics[width=\linewidth]{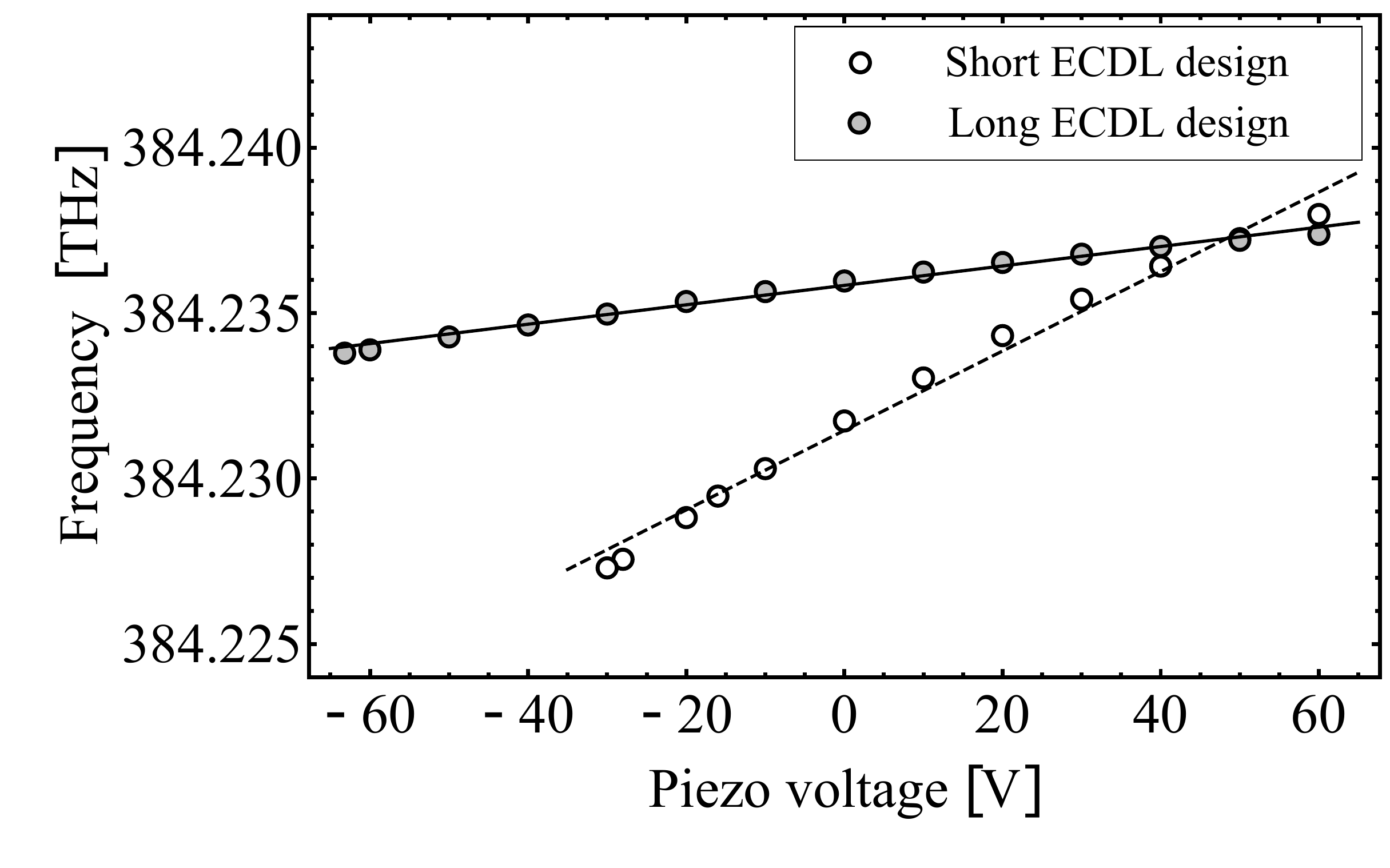}
    \caption{\textbf{Tuning the frequency with a PZT.} Measured laser frequency as a function of the voltage applied to the PZT taken with the short (empty dots) and long (gray dots) ECDL prototypes together with their linear fits. The data were taken by continuously increasing the voltage applied to the PZT. The short cavity design shows a continuous tuning of the laser frequency upon change of the external cavity length in a range of 10.7~GHz (more than two times the FSR) with a slope of 120~MHz/V. After this range it jumps to the next cavity mode. For the long ECDL, the measured range that the PZT can span is 3.6~GHz (about 3~FSR), limited only by the maximum voltage of the supply that we can apply. Note that for both prototypes the mode-hop free range is substantially larger than the FSR of the external cavity. The frequency of the long ECDL varies as 29~MHz/V in this design, with a different slope purely resulting from the different external cavity length. The ratio of the slopes gives a factor of 4.1(2) that agrees with the measured values of FSR.
}\label{Fig:3}
\end{figure}

As shown in the data sets of Fig.~\ref{Fig:3}, we observed a mode-hop free tuning range of 10.7~GHz and 3.6~GHz for the short and long ECDL, respectively. Importantly, these values are in both cases more than twice the FSR of each laser. We note that in the long ECDL case we were limited by the maximum voltage of the power supply, therefore the continuous tuning range can in principle be larger than what measured. We find this large tuning range to be crucially dependent on the alignment of the optical feedback of the grating. 
We attribute this to the fact that if the injection is well optimized, the laser frequency can continuously change for more than a FSR because the laser gain is determined not only by the cavity modes and the grating but also by the photons that are already circulating in the cavity. We note that the mode-hop-free tunability is smaller than the one of a traditional Littrow design in which the grating is rotated around an optimal pivotal point. However, realizing an optimal rotational stage can be challenging, in particular for relatively long ($>$10cm) cavities, in which the rotation of the grating around a non-optimal pivotal point leads to a tunability of a fraction of a FSR [9]. In this case, our design realizes an improvement of the tuning range.
The linear fits of data sets of Fig.\ \ref{Fig:3} give a frequency variation as a function of the voltage applied to the PZT of 120~MHz/V and 29~MHz/V for the short and long ECDL, respectively. The different slopes reflect the fact that the PZT displacements are identical for the two designs but the cavity lengths are different (dL/L $\simeq$ -d$\nu/\nu$). The ratio of the slopes gives a factor of 4.1(2), corresponding to the ratio of the two FSRs and cavity lengths. Deviations from linearity are due to the mechanical properties of the combined system of PZT and rubber o-ring. They can potentially be improved with a different choice of actuators.
Notably, the mode-hop free range of the long cavity laser of 3.6~GHz is a particularly interesting result, since the tuning range of ECDLs with long cavities is usually limited to a small fraction of the FSR due to an un-optimized position of the grating's pivotal point \cite{Kolachevsky:11}.
This results in an increased sensitivity on temperature variations and on the feedback alignment. With our design, instead, it is not necessary to stabilize the external cavity since it is possible to correct for the slow external cavity drifts due to room temperature variations simply by changing the PZT voltage. 
In our prototypes, given the frequency dependence on temperature shown in Fig.\ (\ref{Fig:2}), we can anticipate that temperature changes of up to 0.5$^\circ$C can be compensated for by acting on the piezo elements of the ECDLs.

\subsection{Estimate of the linewidth via heterodyne detection}

\begin{figure}[b!]
	\centering
		\includegraphics[width=\linewidth]{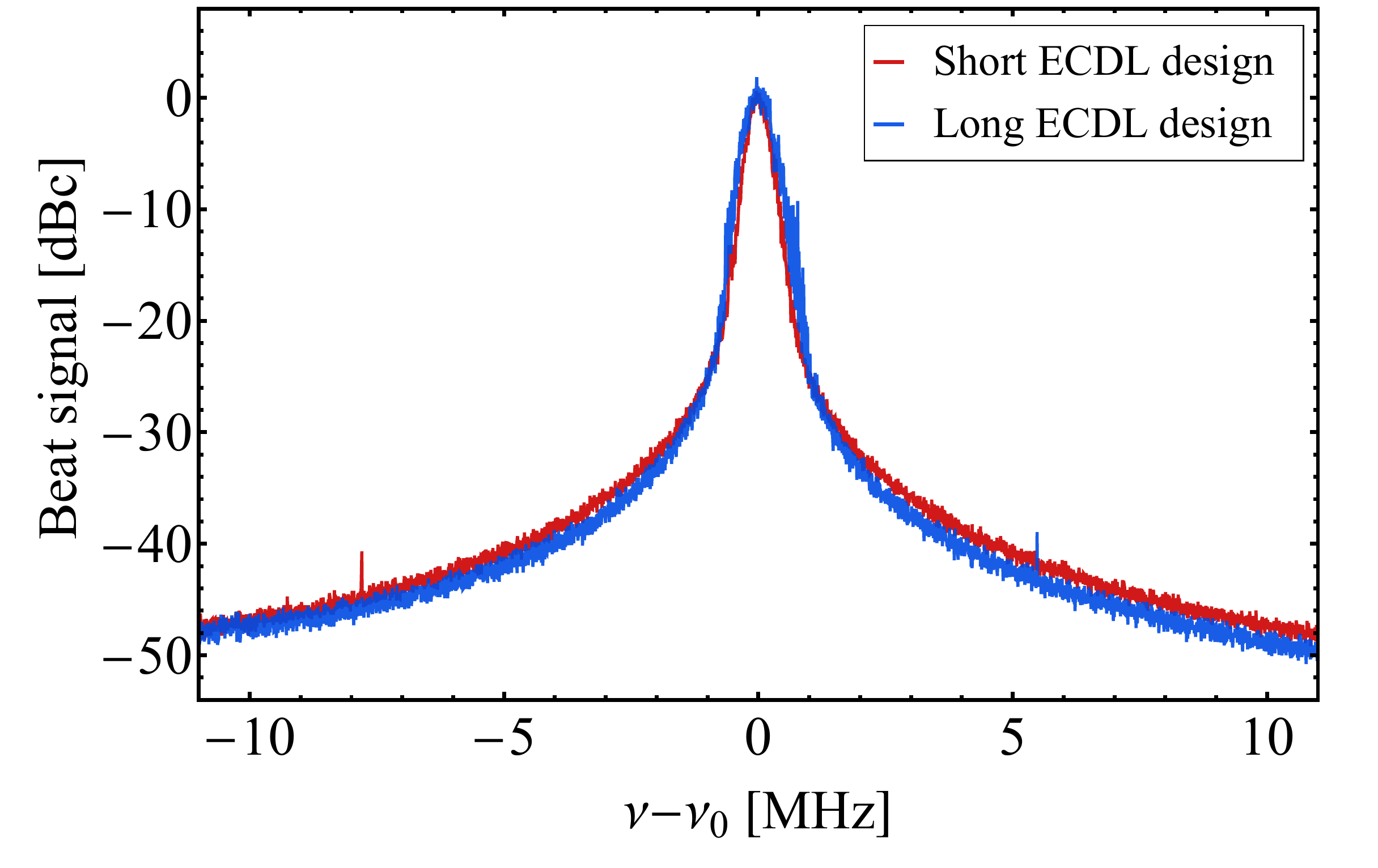}
    \caption{\textbf{Heterodyne beat spectrum.} Photodiode beat signals between the long/short ECDL designs and a Littrow standard ECDL laser (blue/red data) peaked at a frequency $\nu_0$ are recorded by a spectrum analyzer (5~kHz resolution bandwidth, 100 sets average for a total sweep duration of 400~ms). ECDLs are independently locked to Rb lines with a slow feedback to the PZT that does not influence the power spectral density at frequencies $>$~15~Hz. Lorentzian fits of the peaks give a combined FWHM of: 540~kHz (blue data) and 670~kHz (red data), resulting in a linewidth of about 270~kHz (blue data) and 335~kHz (red data) when assuming independent yet identical lasers. 
}\label{Fig:8}
\end{figure}

In order to estimate the linewidth of the laser prototypes and prove that the prototypes can be well stabilized by acting on the PZT even in a noisy environment, we performed a heterodyne beat measurement between each of the new ECDL prototypes and, as a reference, a home-made standard Littrow ECDL with the PZT (Piezomechanik PSt 150/2x3/7) placed behind the grating \cite{RICCI1995}. Each laser is independently locked on the spectroscopy lines of Rubidium by performing standard saturated absorption spectroscopy on a vapor cell. We apply a slow correction (with corner frequency $\sim$15~Hz) to the PZT of all lasers with a PI feedback loop in order to lock the lasers' frequency to the $5 ^2S_{1/2}\,  (F=2) \rightarrow 5^2P_{3/2} \, (F'=2, 3)$ crossover transition in the D$_2$ line of $^{87}$Rb. This atomic reference makes it possible to compensate for slow frequency fluctuations due to temperature drifts of the external cavity and acoustic noise. At frequencies $ >$15 Hz the laser frequency noise is essentially unchanged, therefore the lasers' linewidths are dominated by the remaining acoustic, mechanical and intrinsic noise that are not compensated for by the PZT feedback.
The spectral characteristics of the lasers' emissions are determined from the heterodyne beat signals shown in Fig.\ \ref{Fig:8}. The beat signal carries information about the relative frequency noise of one of the prototype lasers and the reference laser \cite{Nazarathy89}.  By fitting the beat signals with a Lorentzian function, we obtained the combined linewidth for the short and the long ECDLs which are 540~kHz and 670~kHz, resulting in a single laser linewidth of about 270~kHz and 335~kHz, respectively. However, if we assume that the short prototype laser and the reference laser have the same linewidth (i.e.\ 270~kHz), we estimate the linewidth for the long prototype to be 400~kHz. These values agree also with the values of linewidth extracted from of the full widths at $-3$~dB suppression (450~kHz and 700~kHz for the two prototypes) corresponding to individual linewidths of about 225~kHz and 350~kHz for the short and the long ECDLs.
The measured linewidths are in fact comparable to the linewidths of similar designs locked in frequency with just a slow feedback loop \cite{RICCI1995,Kirilov2015}. We attribute the slightly larger width of the long cavity ECDL to the increased mechanical, acoustic and thermal noise of the setup combined with the small bandwidth of the PZT feedback loop which eliminates mostly only the slow drifts due to thermal noise. Consequently, the linewidth reduction is limited since the spectral linewidth is still dominated by acoustical and mechanical noise sources. The linewidth can in principle be further reduced by correcting the higher frequencies on an actuator faster than the PZT, e.g. the laser current \cite{RICCI1995,Kirilov2015} or an intracavity EOM \cite{Stoehr06,Muller06}

\section{Conclusions}

We have realized a new design for a Littrow ECDL in which the small changes of the cavity length are decoupled from the tuning of the grating angle. As a result, the fine adjustment of the laser frequency with the PZT does not affect neither the optical feedback alignment nor the broader grating frequency selection curve, resulting in a better mode-hop stable performance compared to the one of standard Littrow ECDLs without optimized pivotal point. We have assembled two prototypes with different cavity lengths, and observed an improved frequency tunability upon changes of the PZT that allows frequency shifts of more than twice the external cavity free spectral range. This design is particularly suitable for long cavity ECDLs for which the requirement of rotating the diffraction grating around the optimal pivotal point is extremely demanding. Moreover, by using this design it might even be possible to compensate for external cavity temperature drifts only with the backaction of the PZT. 
We also showed that our laser prototype can be frequency locked to an atomic reference by giving a slow feedback signal to the PZT. We compare the laser frequency to a reference laser and measure a linewidth on the order of approx. 300 kHz, a value that is comparable to other ECDL designs. The linewidth could be further improved, if needed, by compensating for high frequency noise with an additional feedback on a faster actuator \cite{RICCI1995,Kirilov2015,Stoehr06,Muller06}.
Finally, the mechanical stability of the ECDL could be further enhanced by placing the laser in a sealed enclosure to cut out noise due to mechanical vibration and pressure changes of the external environment \cite{Cook2012,Kirilov2015}, making it more robust and easy to integrate in e.g. compact optical setups for transportable gravimeters and optical clocks, or space applications.

\paragraph*{Fundings}
European Metrology Programme for Innovation and Research (17FUN07 CC4C); European Research Council (639242); European Research Council (899912).

\paragraph*{Acknowledgements} We thank Massimo Inguscio for continuous support, and Amelia Detti for experimental assistance during the early stage of this work.
The authors declare that part of the results presented in this work are protected by a registered Italian patent (nr. 102019000002013), currently under review for international application  (nr. PCT/IB2020/051055). This work is part of a project that has received funding from the European Research Council (ERC) under the European Union’s Horizon 2020 research and innovation programme (grant agreements no. 899912 and no. 639242). This project 17FUN07 CC4C has received funding from the EMPIR programme co-financed by the Participating States and from the European Union’s Horizon 2020 research and innovation programme.

\paragraph*{Data Availability} Data underlying the results presented in this paper are
available in Ref. \cite{Zenodo}.



\bibliography{ECDLbib}


\end{document}